\documentclass{article}
\usepackage{spconf,amsmath,graphicx}
\usepackage{soul}
\usepackage{hyperref}
\usepackage[dvipsnames]{xcolor} 

\usepackage{amsmath,amsfonts,bm}

\def\eqref#1{equation~\ref{#1}}

\def\1{\bm{1}}

\def\rvx{{\mathbf{x}}}
\def\rvy{{\mathbf{y}}}
\def\rvz{{\mathbf{z}}}

\DeclareMathAlphabet{\mathsfit}{\encodingdefault}{\sfdefault}{m}{sl}
\SetMathAlphabet{\mathsfit}{bold}{\encodingdefault}{\sfdefault}{bx}{n}

\title{Multi-rate adaptive transform coding for video compression}

\twoauthors
 {Lyndon R. Duong\sthanks{Work was performed at Open Codecs, Google LLC.}}
	{
 lyndon.duong@nyu.edu\\
 Center for Neural Science, NYU \\
	 New York, NY, USA}
 {Bohan Li, Cheng Chen, Jingning Han}
	{
 \{bohanli, chengchen, jingning\}@google.com\\
 Open Codecs, Google LLC\\
	Mountain View, CA, USA
	}

\hypersetup{
    colorlinks=true,
    linkcolor=darkgray,
    filecolor=magenta,      
    urlcolor=ForestGreen,
    citecolor=Violet,
    }

\begin{document}
\ninept
\maketitle
\begin{abstract}

Contemporary lossy image and video coding standards rely on transform coding, the process through which pixels are mapped to an alternative representation to facilitate efficient data compression.
Despite impressive performance of end-to-end optimized compression with deep neural networks, the high computational and space demands of these models has prevented them from superseding the relatively simple transform coding found in conventional video codecs.
In this study, we propose learned transforms and entropy coding that may either serve as (non)linear drop-in replacements, or enhancements for linear transforms in existing codecs.
These transforms can be \textit{multi-rate}, allowing a single model to operate along the entire rate-distortion curve.
To demonstrate the utility of our framework, we augmented the DCT with learned quantization matrices and adaptive entropy coding to compress intra-frame AV1 block prediction residuals.
We report substantial BD-rate and perceptual quality improvements over more complex nonlinear transforms at a fraction of the computational cost.
\end{abstract}
\section{Introduction}
\label{sec:intro}
Transform coding is an integral component of image and video coding \cite{goyal2001}.
In state-of-the-art video standards such as HEVC~\cite{sullivan2012overview}, VVC~\cite{bross2021overview}, and AV1 \cite{han2021}, transform coding is used to map block prediction residuals to a domain in which the statistics of the transform coefficients facilitate more effective compression. 
These codecs use linear transforms such as the discrete cosine transform (DCT)~\cite{ahmed1974discrete} and the asymmetric discrete sine transform (ADST)~\cite{han2011jointly}, due to their compression efficiency as well as low computational complexity. 

In recent years, impressive results using end-to-end optimized codecs anticipate a possible shift away from conventional codecs, whose designs are largely based on heuristics and hand-engineered components (\cite{balle2021, yang2021} for reviews).
Indeed, image compression competitions based on rate-distortion (R-D) performance are now dominated by nonlinear machine learning (ML) models \cite{balle2021}.
However, end-to-end ML codecs have yet to become standardized, primarily due to their extreme increase in time and space complexity relative to conventional solutions.
For example, one undesirable factor is that many ML compression approaches train an individual model for each point along the R-D curve, requiring an entirely separate set of neural network parameters for each R-D trade-off. This not only dramatically increases the space needed to store such parameters, but also limits the ability to fine-tune the R-D trade-off. 

\begin{figure}[t]
    \includegraphics[width=1\linewidth]{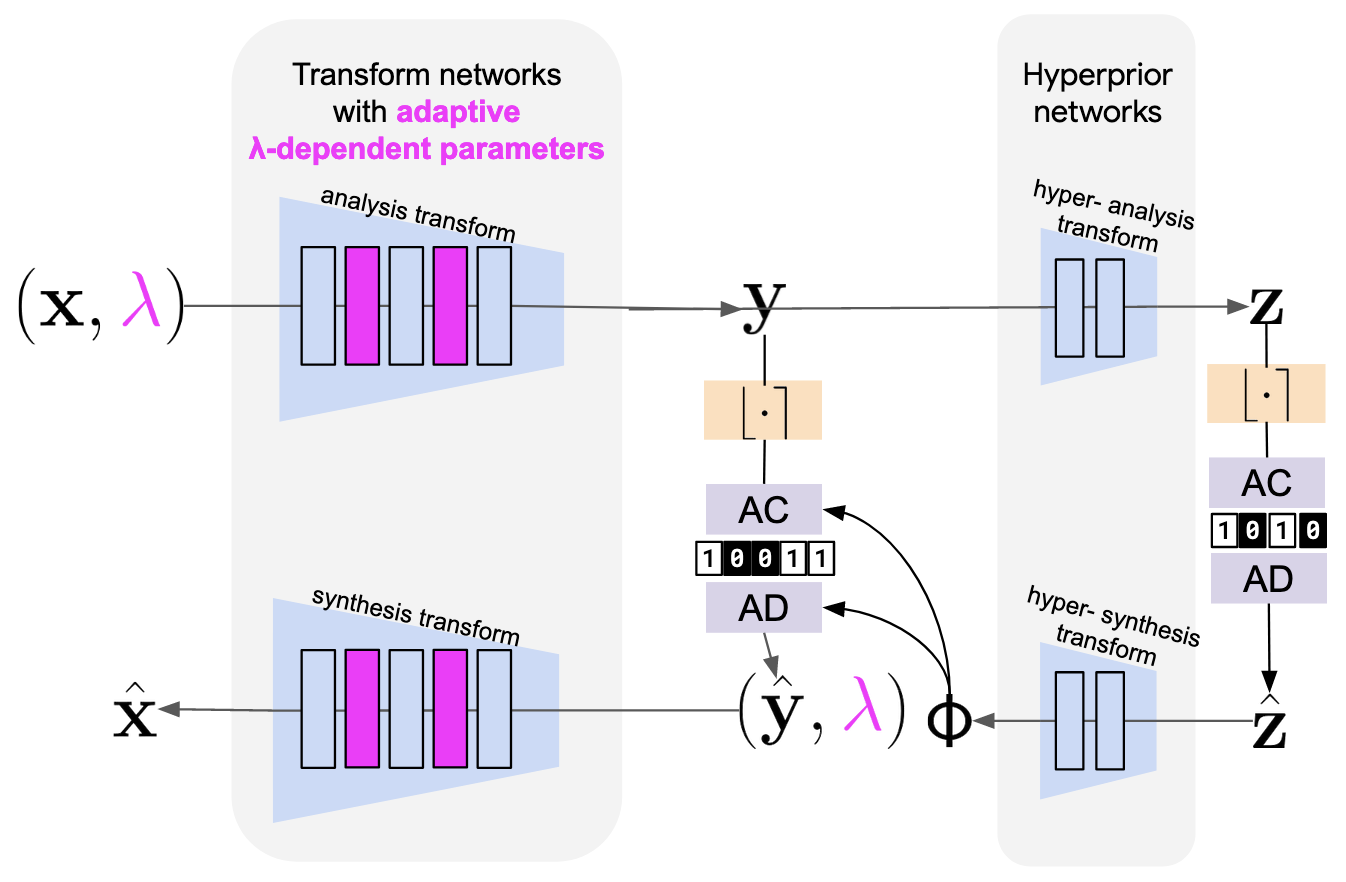}
    \caption{Multi-rate model architecture.
    The analysis/synthesis transforms (left column) can be linear or nonlinear, and use a fixed set of $\lambda$-independent parameters ({\bf \color{RoyalBlue} blue}), but uses a subset of $\lambda$-dependent to fine-tune the R-D trade-off ({\bf \color{Rhodamine} pink}). 
    A learned hyperprior network (right column) enables forward-adaptive entropy coding by conditioning the probability over transform coefficients, $p(\hat{\rvy}\vert \Phi)$. 
    AC and AD denote arithmetic coding and decoding, respectively and
    orange boxes indicate quantization operations.
    See \autoref{sec:methods} for details.
    } 
    \label{fig:arch}
\end{figure}

In this study, we take steps towards addressing these issues, with our contributions summarized as follows:
\begin{enumerate}
    \itemsep0em 
    \item We trained transforms and conditional entropy models with an R-D objective to compress intra-frame prediction residuals collected from the AVM codec~\cite{avm}, the reference software for the next codec from Alliance for Open Media. 
    These models can be used as drop-in replacements, or \textit{augmentations} for existing transform coding modules in video codecs. 
    \item We used a family of architectures and a training procedure that enable multi-rate compression via adaptive gain control with context-adaptive entropy coding~\cite{balle2021}, allowing a single trained model to operate at any arbitrary point along the R-D curve. 
    This vastly improves space complexity compared to training a model for each R-D trade-off.
    \item 
    We augmented the DCT with ML components to provide significant improvements on BD-rate~\cite{bdrate} and structural similarity (SSIM) for intra-frame block prediction residuals compared to learned nonlinear transforms with higher complexity.
\end{enumerate}

\section{Relation to prior work}
\subsection{Transform coding in video codecs}
Conventional video codecs employ inter- and intra-frame prediction to reduce temporal and spatial redundancies.
Subtracting these predictions from the source pixels produces prediction residuals; it is these prediction residuals which are then transformed, quantized, and transmitted~\cite{han2021}.
The purpose of transform coding in lossy compression is to map a signal to a domain in which the components are decorrelated with increased energy compaction.
This representation has redistributed energy that facilitates improved compression.
Our work aligns with other studies in the relatively new field of learned transform coding~\cite{Wang_2022_CVPR, Zhang2020}.
However, rather than learning an end-to-end compression codec using raw pixels as inputs, our models operate on \textit{prediction residuals}. 
Thus, transform coding modules described here can be directly integrated into existing codecs while preserving other specialized codec machinery (e.g. block partitioning, prediction modes, loop filtering, etc.). 

\subsection{Linear transform coding}
In conventional image/video coding, transforms are unitary operations.
For example, in AV1, transform coefficients are obtained via a forward transform, where the transform type is selected from 16 possible 2D-separable linear sinusoidal transforms (e.g. 2D DCT)  through R-D optimization~\cite{han2021}.
A specified quantization index then determines the scales of these transform coefficients in the quantization stage. These quantized transform coefficients are then entropy coded with context-adaptive probability models.
In our work, as an extension to linear transforms, we propose to learn the quantization scalings, and the transform coefficient probability models  directly by using a forward-adaptive scale hyperprior.

\subsection{Nonlinear transform coding}

Work on nonlinear transform coding using machine learning -- typically via neural networks -- is becoming an increasingly popular field of study (\cite{yang2021} for a review).
Arguably the first demonstrably successful neural compression approach was by \cite{balle2016}, where the authors trained a convolutional image autoencoder to optimize an R-D objective.
This approach was done \textit{end-to-end} (i.e. on raw pixels), and included learning the probability model over transform coefficients for the entropy coding step.
In this study, we train our models in a similar fashion to \cite{balle2016,balle2018}; however, rather than training a codec end-to-end, we focus here on the prediction residual transform coding and entropy coding modules.
The multi-rate models with learned context-adaptive entropy coding in our study are similar to those introduced in \cite{balle2021}.
In our work, we propose using this approach as a computationally cost-effective means to augment existing linear transforms in AVM.

\subsection{Conditional entropy models}

There exist two main approaches to adaptive context modeling for entropy models: forward, and backward (i.e. auto-regressive) adaptation \cite{balle2021}.
In AV1, the transform coefficient probability model is updated using backward adaptation \cite{han2021}.
These methods require serial processing of individual transform coefficients to update the probability model.
By contrast, in our work, we use a forward adaptation scheme which uses learned hyper-analysis and hyper-synthesis transforms to transmit side-channel information in order to condition the probability model of the transform coefficients \cite{balle2018}.
One benefit of forward adaptation is that it is amenable to parallel block processing, which shall only become more prevalent as GPUs become more commoditized.

\section{Experimental methods}
\label{sec:methods}

\subsection{Multi-rate compression model architecture}
We present a generalized multi-rate transform and entropy coding framework, with primary focus on multi-rate linear transform coding modules using scale hyperpriors in this study.
The forward-pass of a model is (\autoref{fig:arch})
\begin{align}
     \rvy &\leftarrow g_a(\rvx, \lambda) \label{eq:analysis}\\ 
     \rvz &\leftarrow h_a(\rvy) \label{eq:hyper_analysis}\\ 
     (\mathcal{R}(\hat{\rvz}), \hat{\rvz}) &\leftarrow \texttt{HyperEntropyCoder}(\rvz; \Psi) \label{eq:hyperentropy_coder}\\
     \Phi &\leftarrow h_s(\hat{\rvz}) \label{eq:hyper_synthesis}\\
     (\mathcal{R}(\hat{\rvy}), \hat{\rvy}) &\leftarrow \texttt{EntropyCoder}({\rvy}; \Phi) \label{eq:entropy_coder}\\
     \hat{\rvx} &\leftarrow g_s(\hat{\rvy}, \lambda) \label{eq:synthesis}
\end{align}
where: $\lambda$ is the rate-distortion trade-off parameter; $\rvx$ is a block residual; $\rvy$ is the analysis output; $\rvz$ is the hyper-analysis output; $\hat{\rvy}$ and $\hat{\rvz}$ are the quantized transform and hyper-transform coefficients, respectively; $\Psi$ is a learnable set of probability density parameters for $p(\hat{\rvz}; \Psi)$; $\Phi$, the output of the hyper synthesis transform, is the set of scale parameters for the conditional density $p(\hat{\rvy} \vert \Phi)$; and $\hat{\rvx}$, the output of the synthesis transform, is the block residual reconstruction.

The (hyper-)analysis and (hyper-)synthesis transforms (Equations \ref{eq:analysis}, \ref{eq:hyper_analysis}, \ref{eq:hyper_synthesis}, \ref{eq:synthesis}) are each functions with trainable parameters (e.g. deep neural networks).
The \texttt{HyperEntropyCoder()} uses a probability model over hyper-transform coefficients $p(\hat{z}; \Psi)$, a non-parametric factorized density with learnable parameters $\Psi$ \cite{balle2018}.
Importantly, the hyper-synthesis transform, $h_s(\hat{\rvz})$ outputs the \textit{scale parameters}, $\Phi$, of the transform coefficient probability model to condition the probability density $p(\hat{\rvy} \vert \Phi)$, which is used in the \texttt{EntropyCoder()}.
We chose the conditional density to be a multivariate Gaussian with zero mean.
Entropy coders (Equations \ref{eq:hyperentropy_coder}, \ref{eq:entropy_coder}) output the estimated rates, $\mathcal{R(\cdot)}$ (see \autoref{ssec:obj}) and either dither-quantized (during train time) or quantized coefficients (test time).
See \cite{balle2021} for more details on dithered quantization training.

We performed multi-rate compression using models with a mix of $\lambda$-independent and $\lambda$-dependent parameters (\autoref{fig:arch}; \cite{balle2021, mohan2021}).
The $\lambda$-independent parameters form a universal base model that is shared at each point along the R-D curve.
By contrast, the $\lambda$-dependent parameters dictate where along the R-D curve we wish the model to lie, and serves to \textit{fine-tune} the responses at each layer of the $\lambda$-independent base model.
The $\lambda$-independent parameters are: weights within $g_a$ and $g_s$ (e.g. learned neural network weights, or fixed DCT/inverse DCT); the hyper-transforms, $h_a$ and $h_s$; and the hyperprior parameters, $\Psi$.
The $\lambda$-dependent parameters are element-wise multiplicative scalings on the activations of each layer of $g_a$ and $g_s$ (e.g. learned quantization/dequantization matrices).
Importantly, the hyperprior network does not depend on $\lambda$, and is flexible enough to handle the wide range of outputs of $\rvy=g_a(\rvx, \lambda)$.

\subsection{Learned linear transforms}
We were interested in \textit{extending} the most popular choice of fast linear transforms, the DCT.
The trained linear transforms in this study (Equations \ref{eq:analysis}, \ref{eq:synthesis}) comprised \textit{fixed} linear 2D DCT and inverse DCT, with learned element-wise scalings.
These learned components can be interpreted as (de)quantization matrices, or as simple multiplicative gain modulations on individual transform coefficients.
We studied linear transforms with either the context-adaptive hyperprior entropy model, or a standard factorized entropy model \cite{balle2016}.
We also studied linear multi-rate vs. non-multi-rate (i.e. individual models trained at each $\lambda$) versions of the hyperprior model.

\subsection{Learned nonlinear transforms}

We compared linear models (described above) against learned nonlinear models with factorized entropy models (i.e. no hyperprior) to explore how the coding improvements from augmenting linear models with conditional entropy coding compare to improvements from more complex transforms.
The nonlinear models used in this study were convolutional autoencoders with generalized divisive normalization nonlinear activations after each layer \cite{balle2016}.
Networks were 4 layers deep, with \{256, 128, 64, 64\} square filters with height/width of \{5, 5, 3, 3\}.
Strides were \{2, 2, 1, 1\}, at each layer, respectively.

\subsection{Learned hyperprior transforms and entropy coding}

There were two forms of learned entropy coding used in this study: with and without context adaptation.
Models with adaptation used hyper-analysis ($h_a$) and hyper-synthesis transforms ($h_s$), which were 3-layer multilayer perceptrons (i.e. stacked, fully-connected networks), with \{256, 128, 64\} hidden units in each layer, and rectified linear activations.
Similar to \cite{balle2018}, the \texttt{HyperEntropyCoder} used a non-parametric factorized density with learned parameters and the \texttt{EntropyCoder} used a conditional Gaussian density.
By contrast, models without adaptation lacked hyper-transforms or hyper-entropy coding, and used a learned factorized entropy coder.

\subsection{Stochastic rate-distortion objective}
\label{ssec:obj}

We jointly optimized the parameters of transforms and entropy models by minimizing a Lagrangian rate-distortion loss function with stochastic gradient descent.
For a single sample, $\rvx$, the R-D loss to be minimized is written as
\begin{align}
    \mathcal{L}(\lambda, \rvx, g_a, g_s, h_a, h_s, \Psi) &=   \mathcal{R}(\hat{\rvy}) + \mathcal{R}(\hat{\rvz}) + \lambda \mathcal{D}(\rvx, \hat{\rvx}) \label{eq:objective}.
\end{align}
The first two terms on the RHS are the sum-total rate of $p(\hat{\rvy})$ and $p(\hat{\rvz})$, i.e. the main and side channel transform coefficient densities.
The final term is the distortion between the input block and its synthesized reconstruction, as measured using mean-squared error.

\subsection{Training data and optimization details}
We used AV2 Common Test Conditions (CTC)~\cite{ctc} a3 (720p) and a4 (360p) datasets for training and a2 (1080p) dataset for testing.
Specifically, we obtained luma-channel intra-frame block prediction residuals from AVM \cite{avm}.
In this study, we show results using models trained on prediction residual block sizes of $32 \times 32$.
This procedure yielded approximately $\sim$100K blocks, with 90\% of the data randomly assigned for use during training, and the remaining 10\% reserved for validation.

We minimized \autoref{eq:objective} using stochastic gradient descent with default settings of Adam \cite{kingma2015}. 
Learning rate was 3E-4 for multi-rate, and between [3E-4, 5E-3] for non-multi-rate models.
Batch size was set to 512, and models were trained for 10000 epochs with checkpoints after each epoch.
Results in this study use models corresponding to the epoch with the minimum validation loss. 

Possible values of $\lambda$ were
$\log_2(\lambda) \in \{4, 5, \dots, 17\}$.
Non-multi-rate models were independently trained at each $\lambda$.
For multi-rate training, we randomly sampled $\lambda$ with probability $p(\lambda)=\exp(0.4 i * (N-n)/N)$ where $i$ indexes the possible values of $\lambda$, $n$ is the current and $N$ is the total number of epochs.
Models were implemented using TensorFlow 2 with the TensorFlow Compression library~\cite{tfc_github}, and trained on single NVIDIA P100 or V100 GPUs.





\begin{figure}[ht]
    \centering
    \includegraphics[width=\linewidth]{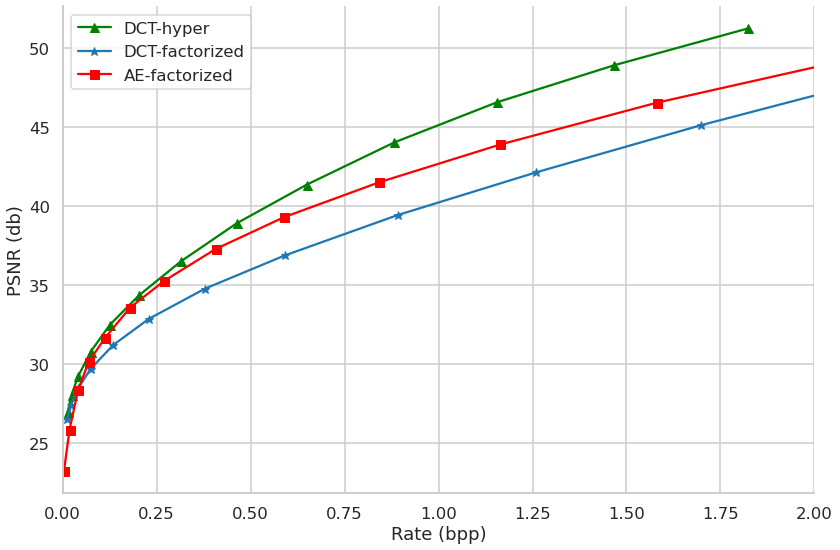}
    \caption{
    DCT with conditional entropy model (DCT-hyper) vs. DCT and nonlinear autoencoder (AE) with factorized entropy model.
    }
    \label{fig:hyper}
\end{figure}

\section{Results}
In this section, we present results demonstrating our method's utility in augmenting fast linear transforms found in AVM.

\subsection{Augmenting linear transforms with forward-adaptive entropy models}

In \autoref{fig:hyper}, we show the coding improvement conferred by including a hyperprior to the linear DCT. 
This allows context adaptation of the transform coefficient probability model at the expense of additional ``side-channel" information via hyper-analysis and hyper-synthesis transforms (\autoref{fig:arch}).
A separate model was trained for \textit{each} R-D trade-off, $\lambda$ (points on \autoref{fig:hyper}).
We also trained nonlinear convolutional autoencoders  without a hyperprior (AE-factorized; i.e. without context adaptation) at each point to evaluate the R-D performance improvements due to the flexibility of the transform compared to the conditional entropy model.
Augmenting the DCT with a learned hyperprior for conditional entropy coding provided a significant improvement in R-D performance (BD-rates: -22.8\% relative to AE-factorized, and -39.3\% relative to DCT-factorized).

\begin{figure}[ht]
    \centering
    \includegraphics[width=\linewidth]{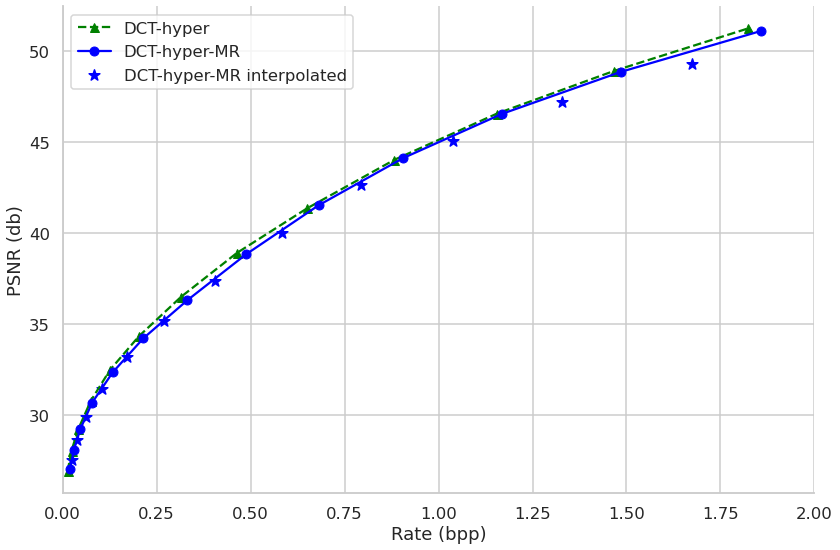}
    \caption{Multi-rate (MR) models closely match the performance of individually-trained networks (which have separately-trained hyperpriors and entropy coders), but can interpolate the R-D curve.}
    \label{fig:rd_interp}
\end{figure}

\subsection{Multi-rate models can interpolate the R-D curve}


Multi-rate training provides a single model that can operate at each point of the R-D curve.
As described in \autoref{sec:methods}, we accomplished this via learned $\lambda$-dependent (de-)quantization matrices, with a jointly learned $\lambda$-independent hyperprior model.
\autoref{fig:rd_interp} shows the R-D performance of a multi-rate model evaluated using values of $\lambda$.

Firstly, \autoref{fig:rd_interp} shows that the multi-rate DCT with hyperprior (DCT-hyper-MR) performed nearly as well as individual, non-multi-rate models trained at each $\lambda$ (DCT-hyper); however, because the multi-rate model used a single $\lambda$-independent hyperprior, it used significantly fewer parameters.
Secondly, by \textit{interpolating} learned $\lambda$-dependent parameters, the
DCT-hyper-MR could compress blocks with \textit{intermediate} values of $\lambda$ and performed (near) optimally despite having never seen these particular values of $\lambda$ at training time.

\begin{figure}[tb]
    \centering
    \includegraphics[width=\linewidth]{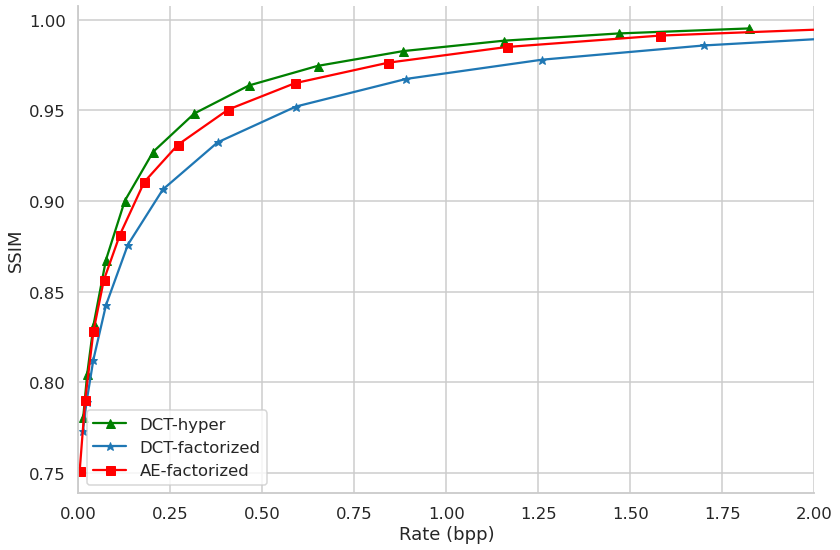}
    \caption{Block reconstruction perceptual quality of DCT with conditional entropy model (DCT-hyper) vs. DCT and nonlinear autoencoder (AE) with factorized entropy models.
    }
    \label{fig:ssim}
\end{figure}

\subsection{Perceptual quality improvements}

There has been growing interest in examining perceptual quality in learned compression~\cite{yang2021}.
We asked how SSIM~\cite{wang2004} of block reconstructions compared for the same models shown in \autoref{fig:hyper}.
Using a more flexible nonlinear model such as a nonlinear autoencoder with factorized entropy model improved SSIM over the DCT-factorized model.
However, AE-factorized uses significantly more parameters than the simple DCT-factorized model where only the DCT (de-)quantization matrices and entropy models were learned ($\mathcal{O}(1\text{E}6)$ vs. $\mathcal{O}(1\text{E}3)$ parameters) .
Interestingly, augmenting the DCT with an adaptive probability model boosted SSIM performance relative to DCT-factorized (-36.4\% BD-rate difference), even further than the SSIM gains observed with AE-factorized (-26.2\% BD-rate difference).
This demonstrates how extending existing linear transforms with relatively simple learned components can provide substantial coding improvements.

\section{Discussion}
\label{sec:discussion}

\subsection{Multi-rate nonlinear transform coding}
In this study, we focused on how our method can be used to \textit{augment} existing fast linear transforms in AVM with learned (de)quantization matrices and hyperpriors.
Our method can also be used to learn more flexible \textit{nonlinear} transforms with hyperpriors.
In preliminary experiments (not shown in this paper), we found that multi-rate nonlinear convolutional architectures (as in \cite{balle2021}) were powerful, but less stable to train than the linear ones presented in this study.
We conjecture that with appropriate tuning of architecture hyperparameters, a multi-rate nonlinear model should outperform the linear multi-rate model in terms of R-D (\autoref{fig:rd_interp}) and SSIM (\autoref{fig:ssim}).
However, this will come at the cost of extreme increase in computational complexity compared with extending existing linear transforms.

Indeed,  since our results revolve around extending and improving linear transforms already found in AVM, differences in computational complexity are primarily derived from the hyperprior $p(\hat{\rvz})$ and conditional probability $p(\hat{\rvy} \vert \hat{\rvz})$ computations.
The hyper-transforms, $h_a$ and $h_s$, were shallow fully-connected neural networks (see Methods), and require $\sim$600 FLOPs/pixel.
By contrast, typical nonlinear autoencoder architectures used in learned compression require $\mathcal{O}(1\text{E}4)$ FLOPs/pixel \cite{mukherjee2022}.
We leave a more formal analysis of computational complexity and comparisons to existing conditional entropy models used in AVM (see below) for follow-up work.

\subsection{Alternative approaches to learned entropy coding}
Conditional entropy models in AV1 and AVM are implemented using backward adaptation, whereby previously decoded transform coefficients are used to inform the context of future encoded coefficients.
This has the drawback of requiring serial processing of individual coefficients.
Conditional entropy coding presented here uses forward adaptation, which is more amenable to parallel processing (e.g. with GPUs).
An important line of future work would be to directly compare the learned transforms and entropy coding presented in this study to existing ones in AVM in order to quantify the advantages of using a data-driven approach.
Recent studies in learned entropy coding have shown improvements through the use of more complex, hybrid forward-backward adaptation schemes (\cite{minnen_joint_2018}; and \cite{yang2021} for a review).
An approach such as this may also yield improvements on the methods used in this study.

\section{Conclusion}
We have proposed a machine learning framework for intra-frame block prediction residual transforms and entropy coding.
These transforms can be multi-rate, and have the ability to interpolate to every point along the R-D curve.
Our method can be used to learn transforms -- either linear or nonlinear -- or augment existing fast linear transforms in existing codecs such as AVM.
Indeed, our results focus on how enhancing the linear DCT used in AVM with learned quantization matrices and context-adaptive entropy models yield significant improvements in R-D performance and SSIM, with relatively low computational cost compared to a learned nonlinear model.
Taken together, our method is a powerful tool that provides practitioners flexibility over varying levels of computational complexity in codec design.

\vfill\pagebreak




\bibliographystyle{IEEEbib}
\bibliography{refs}

\begin{thebibliography}{10}

\bibitem{goyal2001}
V.K. Goyal,
\newblock ``Theoretical foundations of transform coding,''
\newblock {\em IEEE Signal Processing Magazine}, vol. 18, no. 5, pp. 9--21,
  2001.

\bibitem{sullivan2012overview}
Gary~J Sullivan, Jens-Rainer Ohm, Woo-Jin Han, and Thomas Wiegand,
\newblock ``Overview of the high efficiency video coding (hevc) standard,''
\newblock {\em IEEE Transactions on circuits and systems for video technology},
  vol. 22, no. 12, pp. 1649--1668, 2012.

\bibitem{bross2021overview}
Benjamin Bross, Ye-Kui Wang, Yan Ye, Shan Liu, Jianle Chen, Gary~J. Sullivan,
  and Jens-Rainer Ohm,
\newblock ``Overview of the versatile video coding (vvc) standard and its
  applications,''
\newblock {\em IEEE Transactions on Circuits and Systems for Video Technology},
  vol. 31, no. 10, pp. 3736--3764, 2021.

\bibitem{han2021}
Jingning Han, Bohan Li, Debargha Mukherjee, Ching-Han Chiang, Adrian Grange,
  Cheng Chen, Hui Su, Sarah Parker, Sai Deng, Urvang Joshi, Yue Chen, Yunqing
  Wang, Paul Wilkins, Yaowu Xu, and James Bankoski,
\newblock ``A technical overview of av1,''
\newblock {\em Proceedings of the IEEE}, vol. 109, no. 9, pp. 1435--1462, 2021.

\bibitem{ahmed1974discrete}
Nasir Ahmed, T\_ Natarajan, and Kamisetty~R Rao,
\newblock ``Discrete cosine transform,''
\newblock {\em IEEE transactions on Computers}, vol. 100, no. 1, pp. 90--93,
  1974.

\bibitem{han2011jointly}
Jingning Han, Ankur Saxena, Vinay Melkote, and Kenneth Rose,
\newblock ``Jointly optimized spatial prediction and block transform for video
  and image coding,''
\newblock {\em IEEE Transactions on Image Processing}, vol. 21, no. 4, pp.
  1874--1884, 2011.

\bibitem{balle2021}
Johannes Ballé, Philip~A. Chou, David Minnen, Saurabh Singh, Nick Johnston,
  Eirikur Agustsson, Sung~Jin Hwang, and George Toderici,
\newblock ``Nonlinear transform coding,''
\newblock {\em IEEE Journal of Selected Topics in Signal Processing}, vol. 15,
  no. 2, pp. 339--353, 2021.

\bibitem{yang2021}
Yibo Yang, Stephan Mandt, and Lucas Theis,
\newblock ``An introduction to neural data compression,''
\newblock {\em https://arxiv.org/abs/2202.06533}, 2022.

\bibitem{avm}
Alliance for Open~Media,
\newblock ``Aom video model,''
\newblock {\em \\https://gitlab.com/AOMediaCodec/avm}.

\bibitem{bdrate}
Gisle Bjontegaard,
\newblock ``Calculation of average psnr differences between rd-curves,''
\newblock {\em VCEG-M33}, 2001.

\bibitem{Wang_2022_CVPR}
Dezhao Wang, Wenhan Yang, Yueyu Hu, and Jiaying Liu,
\newblock ``Neural data-dependent transform for learned image compression,''
\newblock in {\em Proceedings of the IEEE/CVF Conference on Computer Vision and
  Pattern Recognition (CVPR)}, June 2022, pp. 17379--17388.

\bibitem{Zhang2020}
Xinfeng Zhang, Chao Yang, Xiaoguang Li, Shan Liu, Haitao Yang, Ioannis
  Katsavounidis, Shaw-Min Lei, and C.-C.~Jay Kuo,
\newblock ``Image coding with data-driven transforms: Methodology, performance
  and potential,''
\newblock {\em IEEE Transactions on Image Processing}, vol. 29, pp. 9292--9304,
  2020.

\bibitem{balle2016}
Johannes Ballé, Valero Laparra, and Eero~P. Simoncelli,
\newblock ``Density modeling of images using a generalized normalization
  transformation,''
\newblock 2015.

\bibitem{balle2018}
Johannes Ballé, David Minnen, Saurabh Singh, Sung~Jin Hwang, and Nick
  Johnston,
\newblock ``Variational image compression with a scale hyperprior,''
\newblock 2018.

\bibitem{mohan2021}
Sreyas Mohan, Joshua~L. Vincent, Ramon Manzorro, Peter~A. Crozier, Eero~P.
  Simoncelli, and Carlos Fernandez-Granda,
\newblock ``Adaptive denoising via gaintuning,''
\newblock 2021.

\bibitem{ctc}
``Av2 common test conditions (ctc),''
\newblock {\em \\https:media.xiph.org/video/aomctc/test\_set/}.

\bibitem{kingma2015}
Diederik~P. Kingma and Jimmy Ba,
\newblock ``Adam: A method for stochastic optimization,''
\newblock 2014.

\bibitem{tfc_github}
Johannes Ballé, Sung~Jin Hwang, and Eirikur Agustsson,
\newblock ``{T}ensor{F}low {C}ompression: Learned data compression,''
\newblock {\em http://github.com/tensorflow/compression}, 2022.

\bibitem{wang2004}
Zhou Wang, A.C. Bovik, H.R. Sheikh, and E.P. Simoncelli,
\newblock ``Image quality assessment: from error visibility to structural
  similarity,''
\newblock {\em IEEE Transactions on Image Processing}, vol. 13, no. 4, pp.
  600--612, 2004.

\bibitem{mukherjee2022}
Debargha Mukherjee,
\newblock ``Challenges in incorporating {ML} in a mainstream nextgen video
  codec,''
\newblock {\em Challenge on Learned Image Compression}, 2022.

\bibitem{minnen_joint_2018}
David Minnen, Johannes Ballé, and George~D Toderici,
\newblock ``Joint {Autoregressive} and {Hierarchical} {Priors} for {Learned}
  {Image} {Compression},''
\newblock in {\em Advances in {Neural} {Information} {Processing} {Systems}},
  S.~Bengio, H.~Wallach, H.~Larochelle, K.~Grauman, N.~Cesa-Bianchi, and
  R.~Garnett, Eds. 2018, vol.~31, Curran Associates, Inc.

\end{thebibliography}

\end{document}